\documentclass[a4paper]{jpconf}
\usepackage{graphicx}
\begin{document}
\title{Spin Physics at COMPASS}

\author{Christian Schill on behalf of the COMPASS collaboration}

\address{Physikalisches Institut der Albert-Ludwigs-Universit\"at Freiburg,\\ 
Hermann-Herder Str. 3, 79104 Freiburg, Germany.}

\ead{Christian.Schill@cern.ch}

\begin{abstract} The COMPASS experiment is a fixed target experiment at the CERN SPS using muon and hadron
beams for the investigation of the spin structure of the nucleon and hadron spectroscopy. The main
objective of the muon physics program is the study of the spin of the nucleon in terms of its constituents,
quarks and gluons. COMPASS has accumulated data during $6$ years scattering polarized muons off 
longitudinally or transversely polarized deuteron ($^6$LiD) or proton (NH$_3$) targets.

Results for the gluon polarization are obtained from longitudinal double spin cross section asymmetries
using two different channels, open charm production and high transverse momentum hadron pairs, both
proceeding through the photon-gluon fusion process. Also, the longitudinal spin structure functions of the
proton and the deuteron were measured in parallel as well as the helicity distributions for the three
lightest quark flavours.

With a transversely polarized target, results were obtained with proton and deuteron targets for the
Collins and Sivers asymmetries for charged hadrons as well as for identified kaons and pions. The Collins
asymmetry is sensitive to the transverse spin structure of the nucleon, while the Sivers asymmetry reflects
correlations between the quark transverse momentum and the nucleon spin.

Recently, a new proposal for the COMPASS II experiment was accepted by the CERN SPS which includes
two new topics: Exclusive reactions like DVCS and DVMP using the muon  beam and a hydrogen target to
study generalized parton distributions and Drell-Yan measurements using a pion beam and a polarized
NH$_3$ target to study transverse momentum dependent distributions.
\end{abstract}

\section{Introduction}

COMPASS is a fixed target experiment at the CERN SPS accelerator with a wide physics program focused on the
nucleon spin structure and on hadron spectroscopy. COMPASS investigates the spin structure of the nucleon
in semi-inclusive deep-inelastic scattering. A longitudinally polarized $160$~GeV muon beam is scattered
off a longitudinally  or transversely polarized NH$_3$ or $^6$LiD  target. The scattered muon and the
produced hadrons are detected in a $50$~m long wide-acceptance forward spectrometer with excellent particle
identification capabilities \cite{Experiment}. A variety of tracking detectors are used to cope with the
different requirements of position accuracy and rate capability at different angles. Particle
identification is provided by a large acceptance RICH detector, two  electromagnetic and hadronic
calorimeters, and muon filters. 

The polarized $^6$LiD target is split into two cylindrical cells along the beam
direction. The two cells are polarized in opposite direction. The
polarized NH$_3$ target consists of three cells (upstream, central and
downstream) of 30, 60 and 30 cm length, respectively. The upstream and
downstream cells are polarized in the same direction while the middle cell is
polarized oppositely.

\section{Longitudinal Spin Structure of the Nucleon} The longitudinal spin structure of the nucleon is
investigated by measuring double spin asymmetries in inclusive (DIS) and semi-inclusive (SIDIS)
deep-inelastic scattering on a longitudinally polarized target. SIDIS asymmetries, where pions and kaons
are detected, are sensitive to the individual quark flavours.  The most recent proton asymmetries measured in
2007  \cite{COMPASS-A1, COMPASS-A1-} are shown in figure 1 together with the results from the HERMES
experiment \cite{HERMES-A1, HERMES-A1-}.

\begin{figure}
\begin{center}
\includegraphics[width=0.8\textwidth]{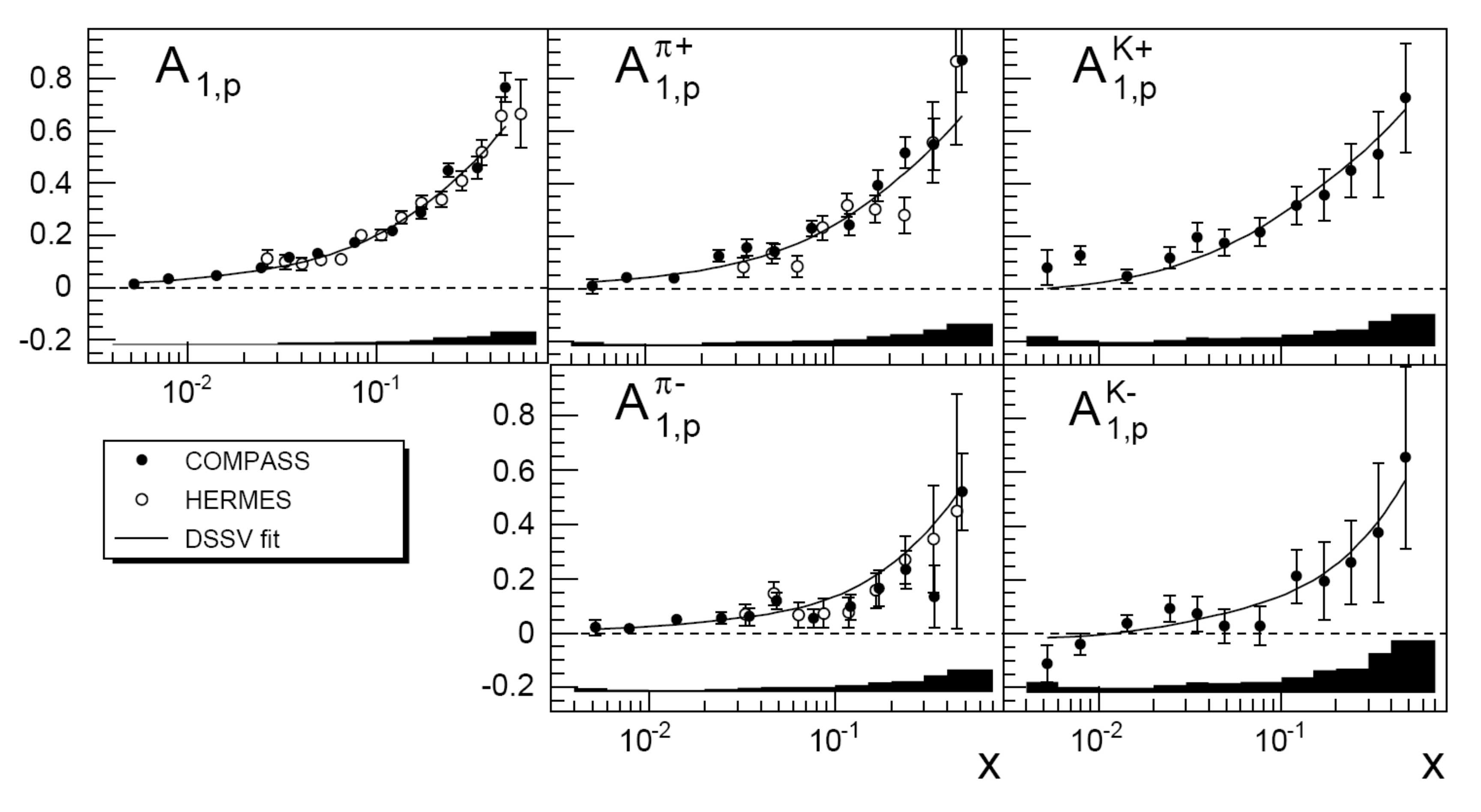}
\end{center}
\label{A1}
\caption{Inclusive, $\pi^\pm$ and $K^\pm$ longitudinal double-spin asymmetries on the proton 
(filled circles: COMPASS results from Ref.~\cite{COMPASS-A1}). For comparison the HERMES results from
 Ref.~\cite{HERMES-A1} (open circles) and the DSSV fit prediction from Ref.~\cite{DSSV} are shown.}
\end{figure}

From the semi-inclusive asymmetries, a leading order QCD extraction of the polarized parton
distribution functions is possible. The measured asymmetries $A_1^h(x, z)$ for a hadron $h$ can be
written as the product of polarized parton distribution functions $\Delta q(x)$ and  fragmentation
functions $D_q^h(z)$, summed over all quark $q$ and anti-quark $\bar{q}$ flavours:

\begin{equation}
A_1^h(x,z)=\frac{\sum_{q,\bar{q}}e^2\cdot \Delta q(x) \cdot D_q^h(z)}
{\sum_{q,\bar{q}}e^2\cdot q(x) \cdot D_q^h(z)}
\end{equation}
The fragmentation functions
are taken from the parametrization of DSS \cite{DSS}. For the unpolarized parton distribution
functions $q(x)$ in the denominator the MRST parametrization \cite{MRST} is used.

In a least-squares fit to the ten inclusive and semi-inclusive asymmetries from proton- and deuteron data,
the quark helicity distributions for $u-$, $d-$ and $s-$quarks are extracted. Since it was observed that in
the  measured $x$-range the extracted strange and anti-strange quark helicity distributions are compatible,
the assumption $\Delta s=\Delta \bar{s}$ has been made to further reduce the number of parameters. The
results of the fit are then shown in figure 2. The curves in 
figures 1 and 2 are the results of a global fit of the DSSV group using previously
existing data at Next-to-Leading Order (NLO) \cite{DSSV,DSSV-}.

\begin{figure}[h]
\includegraphics[width=0.7\textwidth]{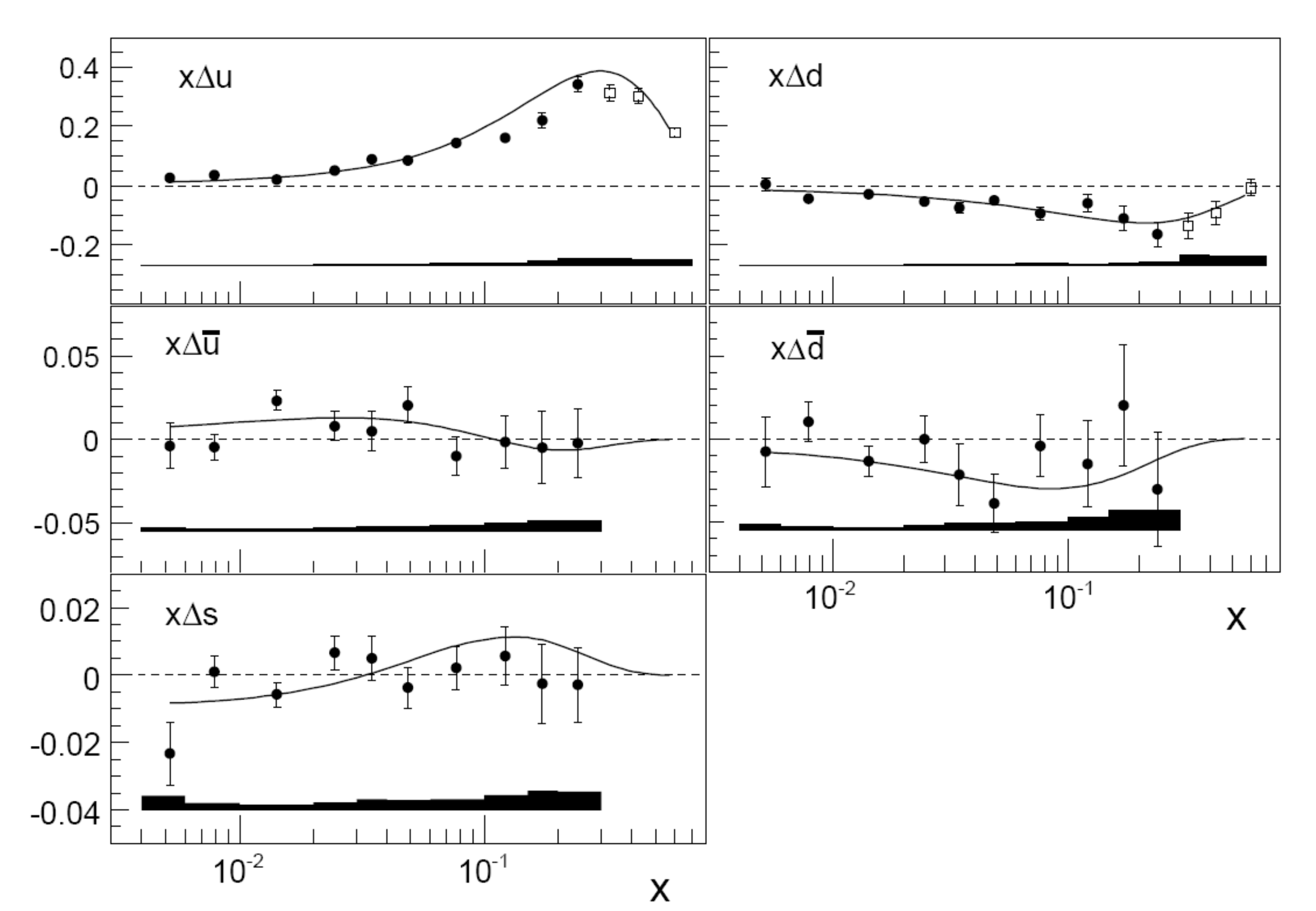}\hspace{2pc}
\label{deltaq}
\begin{minipage}[b]{0.25\textwidth}\caption{Quark helicity distributions $x\Delta u$, $x\Delta d$, $x\Delta \bar{u}$, $x\Delta \bar{d}$
and $x\Delta s$ at $Q^2_0=3$~(GeV/c)$^2$ as a function of $x$. The bands show the systematic
uncertainty of the measurement. The curves are predictions of DSSV calculated at NLO \cite{DSSV}.}
\end{minipage}\end{figure}

The $u$-quark helicity distribution is positive with its maximum in the valence quark region and the
$d$-quark distribution is negative in the same $x$-region. The anti-quark distributions $\Delta
\bar{u}$ and $\Delta \bar{d}$ do not show a significant $x$-dependence. $\Delta\bar{u}$ is
consistent with zero, while $\Delta\bar{d}$ is slightly negative. The flavour asymmetry of the
helicity distribution of the sea, $\Delta \bar{u}-\Delta \bar{d}$ has been extracted from the fit.
The first moment determined in the $x$-range of the experimental data, is found to be 
\begin{equation}
\int_{0.004}^{0.3}(\Delta\bar{u}-\Delta \bar{d})dx=-0.03\pm
0.03(stat.) \pm 0.01 (syst.)
\end{equation}
and therefore compatible with zero. The model prediction $\Delta
\bar{u}-\Delta\bar{d} \simeq \bar{d}-\bar{u}$ of \cite{Peng, Wakamatsu} is not confirmed by the
experimental data. New COMPASS data on a longitudinally polarized proton target have been collected
in $2011$ and will allow the statistical precision of the measurements to be increased.

\section{Gluon polarization in the nucleon} 
At COMPASS, the gluon polarization is directly measured by determining the longitudinal double-spin
asymmetry in the photon-gluon fusion process (PGF). PGF events are searched for in two different channels:
the open charm production channel, where an outgoing charm quark is identified by the production of D$^0$ mesons and
the high-$p_T$ hadron pair channel, where the two outgoing quarks hadronize with high transverse momentum.
The open charm channel provides a clean signature of PGF but with limited statistics. For the high-$p_T$
channel, statistics is large, but the background from other processes is higher.

Open-charm events are selected by reconstructing D$^0$ and D$^*$ mesons from their decay products.
For this analysis, all data collected on $^6$LiD and NH$_3$ from 2002 until 2007 have been used. The
particle identification is performed using the RICH detector. For the D$^*$ sample, the D$^0$ mesons
are tagged by the detection of a low momentum pion from the D$^*\rightarrow $D$^0+\pi_S$ decay. For
untagged events, the decay channel D$^0\rightarrow $K$+\pi$ has been used.  The total number of D$^0$
mesons is about 14.000 (10.000) and 46.000 (19.000) in the D$^*$ and D$^0$ samples collected on $^6$LiD 
(NH$_3$) targets. 

The final result for the gluon polarization from open charm production is given by the weighted mean
of the results for D$^0$ and D$^*$ as \cite{COMPASS-Deltag}

\begin{equation}
\Delta g/g=-0.08\pm 0.21 (stat.)\pm 0.11 (syst.)
\end{equation}
with $<x>=0.11$ at a scale $\mu^2\approx 13$~(GeV/c)$^2$.

\begin{figure}[h]
\includegraphics[width=0.6\textwidth]{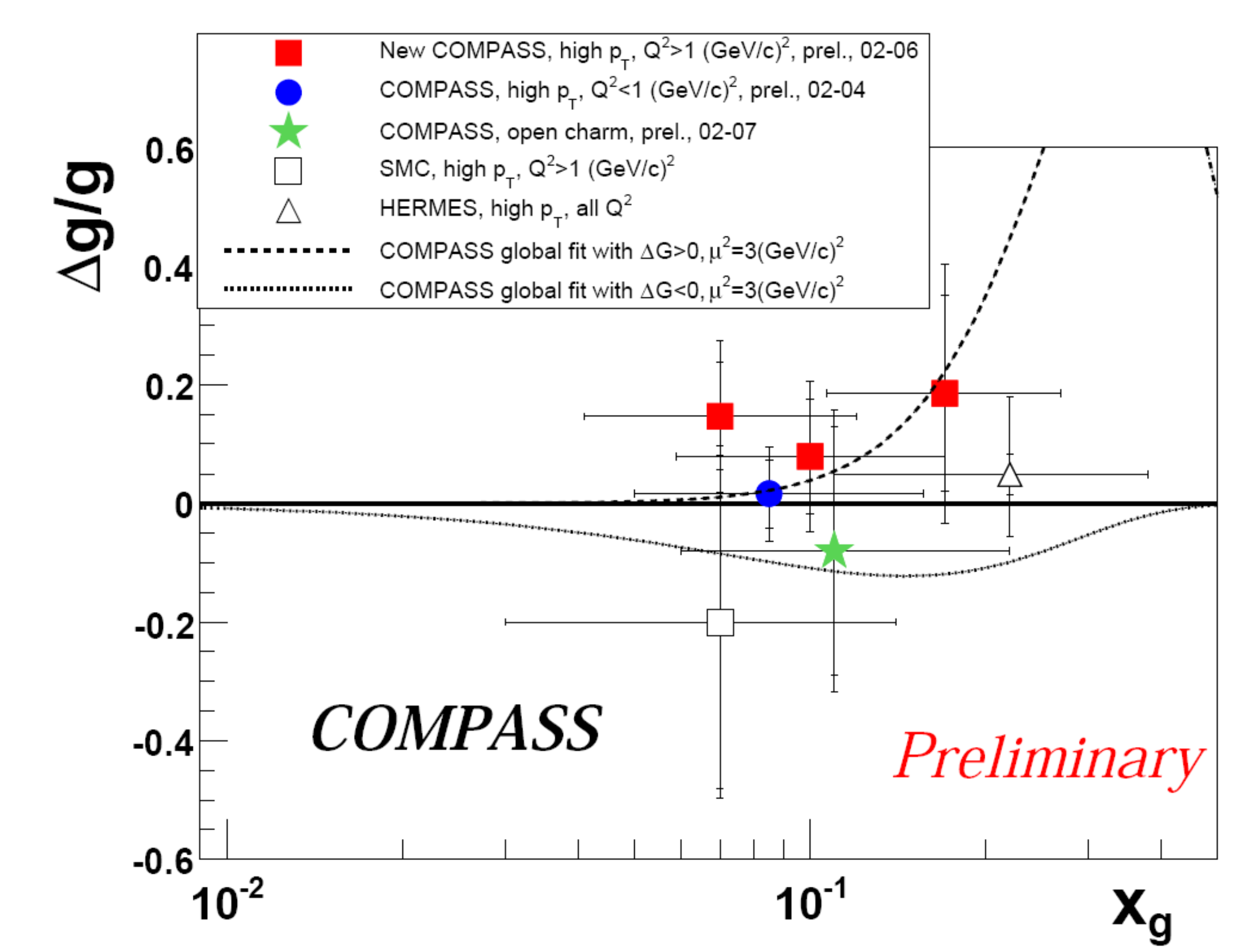}\hspace{2pc}
\label{deltag}
\begin{minipage}[b]{0.3\textwidth}\caption{$\Delta g/g$ measurements from open-charm 
and high-$p_T$ hadron pairs as a function of $x$. The COMPASS results are compared to results of SMC
\cite{SMC} and HERMES \cite{HERMES-DeltaG, HERMES-DeltaG2}. The curves are two  
parameterizations from a COMPASS NLO QCD fit with $\Delta g>0$ (dashed line) and $\Delta g<0$ (solid
line) \cite{QCD-COMPASS}.}
\end{minipage}\end{figure}

In the high-$p_T$ hadron pair channel, other competitive processes contribute to the asymmetry. These are
the leading order process $\gamma + q \rightarrow q$ and the QCD-Compton process $\gamma + q \rightarrow
\gamma + q (g+q)$. Resolved photons are strongly suppressed by selecting events with $Q^2>1$~(GeV/c)$^2$. A
neural network is then used to assign for each event a probability for the three processes (PGF,
QCD-Compton or LO), trained on the results of a Monte Carlo simulation \cite{LEPTO}. The preliminary result
for the gluon polarization for high-$p_T$ hadron pairs with $Q^2>1$(GeV/c)$^2$ is

\begin{equation}
\Delta g/g=0.13\pm0.06 (stat.)\pm 0.06 (syst.)
\end{equation}
at $<x_g>=0.09^{+0.8}_{-0.4}$ at an average scale of $\mu=3$(GeV/c)$^2$. The result has been measured
in three bins of $x_g$ to check for a possible $x_g$ dependence of the gluon polarization. It is
displayed in figure~\ref{deltag} together with the measurements of SMC \cite{SMC} and HERMES
\cite{HERMES-DeltaG, HERMES-DeltaG2} and two parameterizations from a COMPASS NLO QCD analysis of the
world data \cite{QCD-COMPASS}. The  results favor small values of $\Delta g/g$.

\section{Transversity and TMDs} 

Single spin asymmetries in semi-inclusive deep-inelastic scattering (SIDIS) off transversely
polarized nucleon targets have been under intense experimental investigation over the past few years
\cite{COMPASS-T1, COMPASS-T2, COMPASS-T3, COMPASS-T4, HERMES-T}.  They provide new insights into QCD
and the nucleon structure. For instance, they allow the determination of the third 
leading-twist quark distribution function $\Delta _Tq(x)$, the transversity distribution
\cite{Collins,Artru}. Additionally, they give insight into the parton transverse momentum
distribution and angular momentum \cite{Jaffe}.

\begin{figure}[h]
\includegraphics[width=0.75\textwidth]{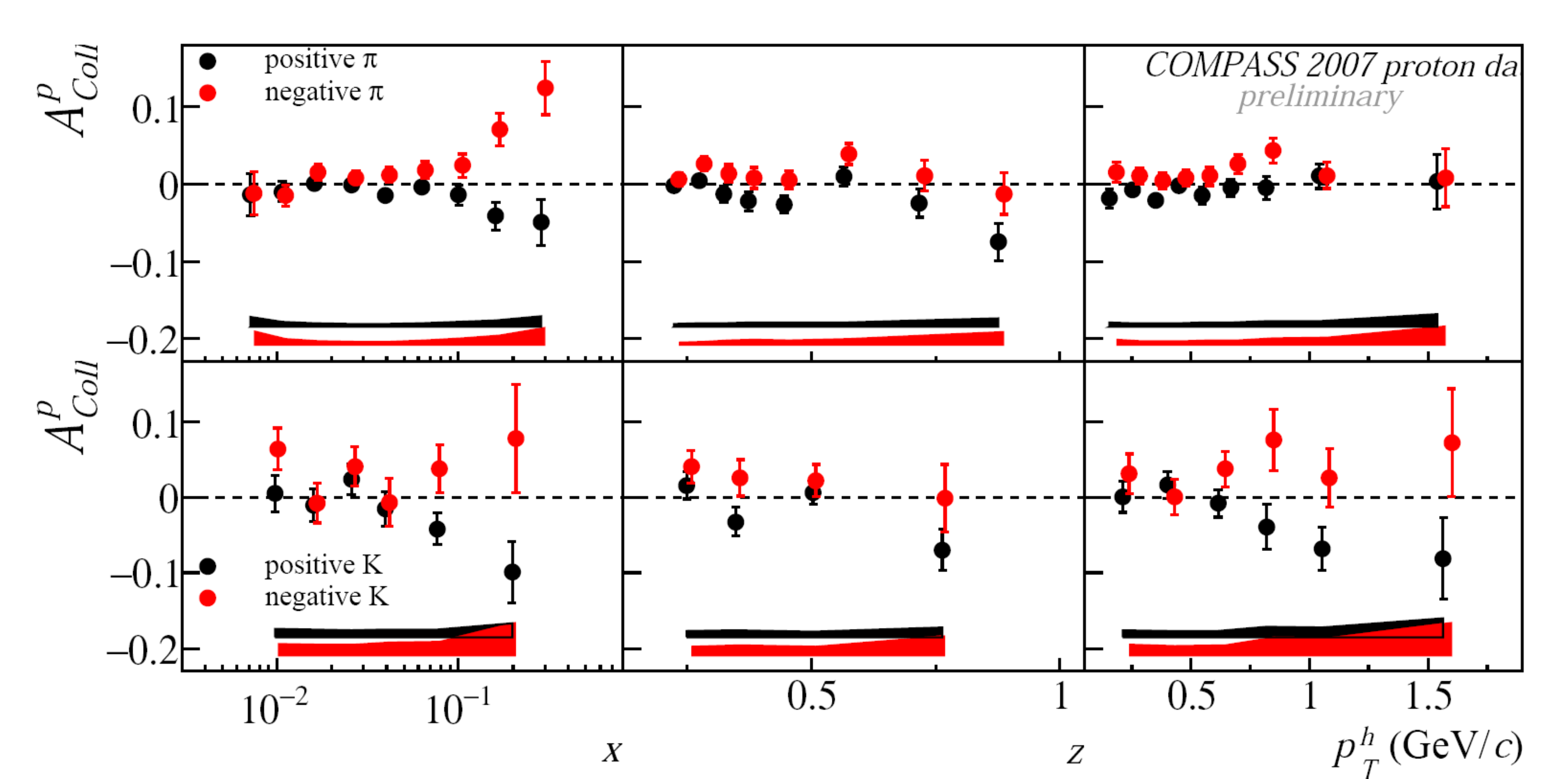}\hspace{1pc}
\label{Collins}
\begin{minipage}[b]{0.20\textwidth}\caption{Collins asymmetry on a proton target for pions (upper row) and kaons (lower
row) as a function of $x$, $z$ and $p_T$ \cite{COMPASS-T1}. The bands
show the systematic uncertainty of the measurement.}
\end{minipage}\end{figure}

The measurement of transverse spin effects in semi-inclusive deep-inelastic scattering is an
important part of the COMPASS physics program. In part of the years 2002-2004 data were taken by
scattering a $160$~GeV muon beam on a transversely polarized deuteron target. In 2007 and 2010,
additional data were collected on a transversely polarized proton target.

In semi-inclusive deep-inelastic scattering the transversity distribution $\Delta_Tq(x)$ can be
measured in combination with the chiral odd Collins fragmentation function. According to Collins,
the fragmentation of a transversely polarized quark into an unpolarized hadron generates an
azimuthal modulation of the hadron distribution  with respect to the lepton scattering plane
\cite{Collins}. 

In Fig.~\ref{Collins} the results for the Collins asymmetry on the proton target are shown as a function
of $x$, $z$, and $p_T$ for positive and negative  hadrons \cite{COMPASS-T1}. For
small $x$ up to $x=0.05$ the measured asymmetry is small and statistically compatible with zero, while
for the last points an asymmetry different from zero is visible. The asymmetry increases up to about
$8$\% with opposite sign for negative and positive hadrons. This result confirms the measurement of a
sizable Collins function and transversity distribution.

COMPASS has measured the Collins asymmetries on a deuteron target as well. They are all compatible with
zero \cite{COMPASS-T4}.  From this measurement the opposite sign of $u$- and $d$- quark  transversity has
been derived. Both, proton and deuteron data sets have been employed in global fits taking into account the
Collins fragmentation function from BELLE and the Collins asymmetries from COMPASS and HERMES to obtain
constrains to the transversity distribution for $u$- and $d$-quarks \cite{Bacchetta}.

Another azimuthal asymmetry is related to the Sivers effect. The Sivers asymmetry arises from a coupling of
the intrinsic transverse momentum $\overrightarrow{k}_T$ of unpolarized quarks with the spin of a
transversely polarized nucleon \cite{Sivers}. The correlation between the transverse nucleon spin and the
transverse quark momentum is described by the Sivers distribution function  $\Delta_0^Tq(x,
\overrightarrow{k}_T)$. Since the Collins and Sivers asymmetries are independent azimuthal modulations of
the cross section for semi-inclusive deep-inelastic scattering, both asymmetries are determined
experimentally from the same dataset.

In Fig.~5 the results for the Sivers asymmetry on the proton are shown as a function of $x$,
$z$, and $p_T$.  The Sivers asymmetry for negative hadrons  is small and statistically compatible with
zero. For  positive hadrons the Sivers asymmetry is positive \cite{COMPASS-T1}.  The Sivers asymmetry
on the deuteron target  is small and compatible with zero, which is due to the opposite sign of the $u-$
and $d-$quark Sivers function \cite{COMPASS-T4}.

\begin{figure}[h]
\includegraphics[width=0.75\textwidth]{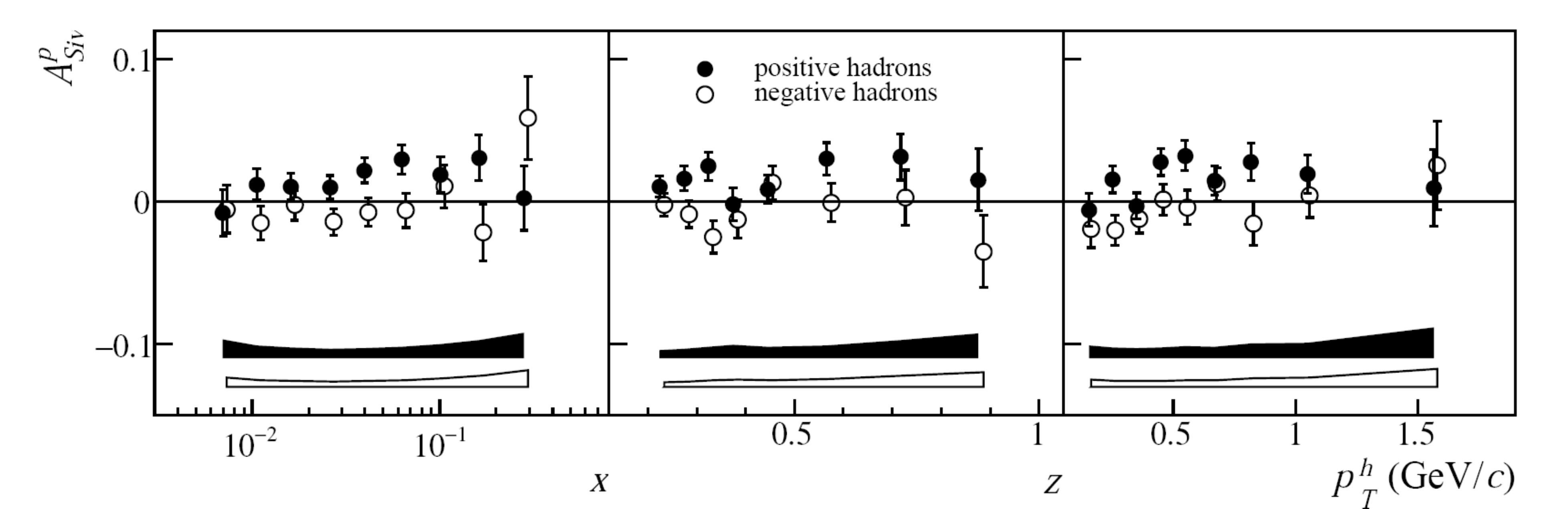}\hspace{1pc}
\label{Sivers}
\begin{minipage}[b]{0.20\textwidth}\caption{Sivers asymmetry on a proton target
for positive and negative hadrons as a function of $x$, $z$ and $p_T$ from \cite{COMPASS-T1}.}
\end{minipage}\end{figure}

\newpage
\section{Outlook}
Recently, the COMPASS II  proposal \cite{COMPASSII} was submitted to improve the knowledge of the
momentum structure of the nucleon towards a three dimensional picture. For this a series of new
measurements is planned. A study of generalized parton distributions (GPD) will be done using exclusive
reactions like deeply virtual Compton scattering (DVCS) and deeply virtual meson production (DVMP)
\cite{DVCS,DVCS2}. Drell-Yan processes will be used for a complementary study of transverse momentum dependent
distributions (TMD) using a transversely polarized target \cite{TMD}. At very low momentum transfers
Primakoff reactions can be used to extract pion and kaon polarizabilities.  The COMPASS II proposal was
approved on December 2010 for an initial data taking of three years.

\section*{Acknowledgments}
This work has been supported in part by the German BMBF.

\end{document}